# Giant intrinsic rectification and nonlinear Hall effect under time-reversal symmetry in a trigonal superconductor

Yuki M. Itahashi[1], Toshiya Ideue[1]*, Shintaro Hoshino[2], Chihiro Goto[1], H. Namiki[3], T. Sasagawa[3], Yoshihiro Iwasa[1,4]

[1] *Quantum-Phase Electronics Center (QPEC) and Department of Applied Physics, The University of Tokyo, Tokyo 113-8656, Japan*

[2] *Department of Physics, Saitama University, Saitama 338-8570, Japan*

[3] *Laboratory for Materials and Structures, Tokyo Institute of Technology, Kanagawa 226-8503, Japan*

[4] *RIKEN Center for Emergent Matter Science (CEMS), Wako 351-0198, Japan*

*Corresponding author: ideue@ap.t.u-tokyo.ac.jp

**Abstract**

**Nonreciprocal or nonlinear responses in symmetry-broken systems are powerful probes of emergent properties in quantum materials, including superconductors, magnets, and topological materials. Recently, vortex matter has been recognized as a key ingredient of giant nonlinear responses in superconductors with broken inversion symmetry. However, nonlinear effects have been probed as excess voltage only under broken time-reversal symmetry. In this study, we report intrinsic rectification and nonlinear anomalous Hall effect under time-reversal symmetry in the noncentrosymmetric trigonal superconductor $PbTaSe_2$. The magnitude of anomalous nonlinear transport is two orders of magnitude larger than those in the normal state, and the directional dependence of nonlinear signals**

**are fully consistent with crystal symmetry. The enhanced nonlinearity is semiquantitatively explained by the asymmetric Hall effect of vortex-antivortex string pairs in noncentrosymmetric systems. This study enriches the literature on nonlinear phenomena by revealing a novel aspect of quantum transport in noncentrosymmetric superconductors.**

Recently, symmetry breaking in solids has become the focus of research in condensed matter physics. It is also a key strategy for developing novel functionalities. To date, many characteristic physical properties, which are unique to noncentrosymmetric crystals, have been reported. For example, the nonlinear optical response such as the second-harmonic generation and optical parametric effect are known to occur in noncentrosymmetric crystals. Broken inversion symmetry also affects transport properties via asymmetric scattering, spin–orbit interaction, magnetic structure, and accompanying geometrical/topological characteristics.

Among the various emergent transports originating from symmetry breaking, the second-order nonlinear transport, which includes the intrinsic rectification effect and nonlinear Hall effect, is recognized as a sophisticated probe of symmetry breaking and a potential functionality for rectifying a variety of quantum currents[1-21]. To date, it has been studied mainly in systems with broken time-reversal symmetry[1–14]. Recently, however, it has been proposed that second-order nonlinear transport can occur even under time-reversal symmetric conditions. An important example is the nonlinear anomalous Hall effect[18], which is a new type of Hall effect realized under time-reversal symmetric conditions and has been experimentally observed in few-layer $WTe_2$[19,20] and bulk $TaIrTe_4$[21]. Band topology/geometry (i.e., Berry curvature dipole)[19] and anomalous scattering (skew-scattering-like mechanism)[20] have been reported to be the origins of the nonlinear Hall effect. However, no occurrence of the intrinsic rectification effect in the longitudinal resistance under time-reversal symmetric conditions has been reported so far, and it is necessary to investigate more materials that show the anomalous nonlinear Hall effect with a distinctive origin. Furthermore, the search for anomalous nonlinear transport in exotic quantum states, such as superconductivity, is a significant challenge from both fundamental and technological points of view.

In this study, we investigated the second-order nonlinear anomalous transport in PbTaSe$_2$, which is a noncentrosymmetric trigonal superconductor and has attracted increased interest as a possible topological material[22–27]. We observed a nonlinear anomalous Hall effect, as well as a rectification effect that satisfies the characteristic directional dependence for the trigonal symmetry. Remarkably, we found that the nonlinear anomalous transport under the time-reversal symmetric condition was enhanced by orders of magnitude in the superconducting (SC) state. Second-order nonlinear transport exhibited a peak structure around the superconducting transition, indicating that vortices/antivortices excited in the layered material played a major role in nonlinear transport. The observed superconducting nonlinear Hall effect/rectification effect can be explained by the asymmetric Hall effect of vortices and antivortices owing to the rectification by trigonal potentials.

PbTaSe$_2$ is composed of alternating stacking of 1H-TaSe$_2$ and Pb layers (Fig. 1). Because each 1H-TaSe$_2$ layer has a noncentrosymmetric trigonal structure (Fig. 1a) and the stacking direction is the same for all the TaSe$_2$ layers (Fig. 1b), multilayer PbTaSe$_2$ also has a noncentrosymmetric trigonal symmetry[22–24]. Furthermore, it exhibits a superconducting transition at $T_c$ = 3.7–3.8 K[22,23]; therefore, it is an ideal platform for investigating second-order nonlinear transport in superconducting states.

Thin flake samples are highly beneficial for the study of nonlinear transport because a large current density is easily obtained. Therefore, we fabricated micro-size PbTaSe$_2$ devices (Fig. 1c) with typical thicknesses of approximately 100 nm through the exfoliation method. We prepared samples with two different configurations (Fig. 2a): the current flowing along a zigzag direction (configuration A: samples 1, 3, and 6) and armchair direction (configuration B: samples 2, 4, 5, and 7). The crystal orientation was well identified by the flake shapes[28] and scanning transmission electron microscopy (STEM) measurements. For example, in Fig. 1d, we depict the STEM image of the cross-section along the black dashed line in Fig. 1c. The STEM image is identical to the cross-sectional image of PbTaSe$_2$ along the armchair direction (inset in Fig. 1d), indicating that the current flows in the zigzag direction in sample 1. Fig. 1e depicts the temperature variations of first harmonic resistance $R_{xx}^{\omega}$ in sample 2. It indicates a metallic behavior with a residual resistivity ratio value of approximately 100, which is consistent with the values reported in previous studies on bulk crystals[23–25]. Figure 1f depicts a magnified view of the temperature dependence of $R_{xx}^{\omega}$ around the superconducting transition. The superconducting transition temperature, $T_c$, defined as the midpoint of the resistive transition ($B = 0$ T, red line) is 3.6 K, which is also consistent with previous studies on bulk crystals[22,23]. When a magnetic field of 1 T was applied (orange line), the superconducting transition was completely suppressed. In the following, we discuss transport only under the time-reversal symmetric condition, that is, without a magnetic field, unless stated otherwise.

First, we focus on the second-harmonic resistance in the normal state. In trigonal crystals, a second-order nonlinear voltage can appear along the armchair direction when current is applied along either the armchair or zigzag direction (see Methods). The intrinsic rectification effect (nonlinear anomalous Hall effect) is expected in response to the applied current along the armchair (zigzag) direction (Fig. 2a). A recent theory proposed second-order nonlinear transport in trigonal systems under time-reversal symmetry[29] as well as a possible mechanism. Note that a trigonal crystal has three mirror planes and is thus nonpolar. This is in contrast to WTe$_2$ with only one mirror plane, along which the second-order nonlinear transverse voltage has been observed[19,20]. In trigonal crystals, the effect of the Berry curvature dipole can be eliminated owing to the high symmetry. Therefore, we can investigate other possible origins such as skew scattering or unknown effects such as a higher-order Berry curvature distribution in momentum space.

Figures 2b and 2c depict the current dependences of second harmonic resistance ($R^{2\omega}$) in configuration A (sample 3, $T = 20$ K) and configuration B (sample 4, $T = 50$ K), respectively. When $I$ is applied along the zigzag (armchair) direction, the finite $I$-linear $R_{yx}^{2\omega}$ ($R_{xx}^{2\omega}$) is observed, which is significantly larger than $R_{xx}^{2\omega}$ ($R_{yx}^{2\omega}$). These results are consistent with the above symmetry argument for the threefold rotational symmetry (see Methods), which unambiguously excludes the possibility of unexpected nonlinear responses coming from extrinsic effects such as asymmetric shapes and/or configurations of electrodes. Note that the intrinsic rectification effect (Fig. 2c), which has never been reported under time-reversal symmetry, is clearly observed as well as the nonlinear anomalous Hall effect (Fig. 2b).

To further understand the nonlinear anomalous Hall effect in the normal state, we measured the temperature variations of the normalized nonlinear anomalous Hall signal. $\frac{|E_y^{(2)}|}{\left(E_x^{(1)}\right)^2}$ (see Methods) and linear conductivity $\sigma_{xx}^\omega = \left(\frac{Wt}{L} R_{xx}^\omega\right)^{-1}$ of sample 1 are depicted in Fig. 2(d). Here, $W = 3.5$ μm, $L = 1.7$ μm, and $t = 123$ nm are the sample width, distance between electrodes, and sample thickness, respectively. Both $\frac{|E_y^{(2)}|}{\left(E_x^{(1)}\right)^2}$ and $\sigma_{xx}^\omega$ increased monotonically with a decrease in the temperature, exhibiting similar behavior. In Fig. 2(d), we analyze the correlation between these quantities by plotting $\frac{|E_y^{(2)}|}{\left(E_x^{(1)}\right)^2}$ versus $(\sigma_{xx}^\omega)^2$. $\frac{|E_y^{(2)}|}{\left(E_x^{(1)}\right)^2}$ indicates a linear dependence on $(\sigma_{xx}^\omega)^2$, particularly in the high $\sigma_{xx}^\omega$ (low-temperature) region. Unexpectedly, a nonlinear anomalous response was visible even above $T = 100$ K. In general, the second-order nonlinear transverse voltage can be well described by equation $\frac{|E_y^{(2)}|}{\left(E_x^{(1)}\right)^2} = \xi(\sigma_{xx}^\omega)^2 + \eta$ (where $\xi$ and $\eta$ are phenomenological fitting parameters)[21], reflecting the two contributions to the nonlinear anomalous Hall effect: the first term can be the skew-scattering-like origin, which scales as $\frac{|E_y^{(2)}|}{\left(E_x^{(1)}\right)^2} \propto \tau^2 \propto (\sigma_{xx}^\omega)^2$, and the second term that satisfies $\frac{|E_y^{(2)}|}{\left(E_x^{(1)}\right)^2} \propto \tau^0 \propto (\sigma_{xx}^\omega)^0$ corresponds to scattering-free mechanisms such as the Berry curvature dipole effect and the side-jump mechanism. In sample 1, fitting parameters $\xi$ and $\eta$ are estimated as $\xi = 2.3 \times 10^{-20}$ m$^3$V$^{-1}\Omega^2$ and $\eta = -3.2$ μmV$^{-1}$, respectively. Unlike a previous study on WTe$_2$[20], in which both contributions cannot be neglected, the first term is dominant in PbTaSe$_2$. This might be because the Berry curvature dipole will strictly vanish in trigonal crystals with three mirror planes. Note that a small deviation from the relation $\frac{|E_y^{(2)}|}{\left(E_x^{(1)}\right)^2} \propto \tau^2 \propto (\sigma_{xx}^\omega)^2$ (black linear dashed line in Fig. 2(e)) was observed in the low $\sigma_{xx}^\omega$ (high-temperature) region. This might be attributed to the contribution from the $\sigma_{xx}^\omega$-linear term ($\frac{|E_y^{(2)}|}{\left(E_x^{(1)}\right)^2} \propto \sigma_{xx}^\omega$) originating from both skew and side-jump scatterings[30]. The same scaling of $\frac{|E_y^{(2)}|}{\left(E_x^{(1)}\right)^2}$ and $(\sigma_{xx}^\omega)^2$ was observed in other samples, as depicted for sample 3 in Supplementary Fig. S5.

Next, we focus on nonlinear transport in the SC state. Figures 3a and 3b depict the current dependences of $R^{2\omega}$ (left) and $R_{xx}^\omega$ (right) in configurations A (sample 1) and B (sample 2), respectively, at $T = 2$ K. With an increase in the current, the superconducting zero-resistance state was broken and a finite resistance state appeared (black dotted curve). Around this transition, a sharp peak of $R_{yx}^{2\omega}$ ($R_{xx}^{2\omega}$) was observed when $I$ was applied parallel to the zigzag (armchair) direction. Note that such anomalies are negligibly small in other directions, in fair agreement with the directional dependence of second-order nonlinear transport in the trigonal systems, as in the case of the normal state (Figs. 2b and 2c). Figures 3c and 3d depict the temperature dependences of $R^{2\omega}$ (left) and $R_{xx}^{2\omega}$ (right) in configurations A (sample 1, $I = 0.06$ mA) and B (sample 2, $I = 0.3$ mA), respectively. A peak behavior similar to Figs. 2a and 2b was observed in $R_{yx}^{2\omega}$ ($R_{xx}^{2\omega}$), whereas such a signal was small or absent in the other direction when *I* was parallel to the zigzag (armchair) direction. Figures 3a–d indicate that both the nonlinear anomalous Hall effect and the rectification effect were significantly enhanced in the transition region and suppressed in the zero-resistance state. Such nonlinear anomalous transport, which satisfies the directional dependence of trigonal symmetry and is enhanced in the SC fluctuation region, was observed in all the samples we measured (Supplementary Materials V).

Around the superconducting transition, excited vortex–antivortex pairs or vortex loops are known to cause a resistive state in 2D or layered superconductors even under the time-reversal symmetric condition[31,32]. In the present case of layered PbTaSe$_2$, our simulation revealed that the vortex–antivortex string pair had the lowest energy excitation, as described in Supplementary Material (Section III). Therefore, the system can be regarded as 2D from the vortex point of view. We propose that this vortex/antivortex dynamics causes the nonlinear anomalous Hall effect during the superconducting transition, as discussed below, in a manner similar to the vortex rectification effect in trigonal superconductors under an out-of-plane magnetic field[12,15].

In Fig. 3e, we depict a possible mechanism for the observed nonlinear anomalous Hall effect in trigonal superconductors by considering the asymmetric vortex/antivortex Hall effect owing to the trigonal potential. The first clue came from the observation of the excess component in the Hall resistance, which was interpreted as a vortex Hall effect[33–40] (see Supplementary Material Section II, Supplementary Figs. S1 and S2). The origin of the vortex Hall effect is still being debated. One potential mechanism is the charging of the vortex core owing to the difference between the chemical potentials of the normal core and superconducting states. We consider that the vortices and antivortices are excited by a finite temperature or current as string pairs even without magnetic fields (see Supplementary

Material Section III, Supplementary Fig. S3). When current is applied along the zigzag direction (configuration A), vortices/antivortices are first driven in the armchair direction and then curved in the transverse zigzag direction owing to the vortex/antivortex Hall effect. During this process, vortices/antivortices are rectified, reflecting the trigonal potential; therefore, the vortex Hall effect is asymmetric. This results in the antiparallel motion of vortices and antivortices, which is equivalent to the net flow of vorticity current (purple arrow) in Fig. 3e; the excess voltage appears perpendicular to it, or along the armchair direction, and is observed as the nonlinear anomalous Hall voltage. A similar scenario also explains the intrinsic rectification effect (see Supplementary Material, Supplementary Fig. S3).

This model is formulated in Supplemental Material, Section IV. In this theoretical description, we consider the vortex/antivortex dynamics, particularly the Hall effect in trigonal potentials (Supplementary equation (S2)). By combining the rectification effect and Hall effect of vortices/antivortices[15], we obtained the expression of $R_{yx}^{2\omega}$ as $R_{yx}^{2\omega} = \frac{(\phi_0^*)^3 n_v r \ell_v I}{k_B T W \eta_0} g_2\left(\frac{U}{k_B T}\right)$, where $\phi_0^*$ is the flux quanta, $n_v$ is the total number density of the vortices/antivortices, $r$ is the Hall angle of vortices, and $\eta_0$ is the friction coefficient. Parameters $\ell_v, U,$ and $g_2$ are dimensionless length, energy, and function, respectively, and are determined from the details of the potential shape of the vortices. In this study, we assumed that the dissociation of vortex–antivortex pairs is induced predominantly by the current injected for the observation of the nonlinear transport effect. Employing the realistic values of the phenomenological parameters, we estimated the value of $R_{yx}^{2\omega}$ to be approximately 1.4 m$\Omega$ near the superconducting transition temperature (Supplementary Fig. S4b), which is in good agreement with the experimental results. Our theoretical model can also explain the temperature dependence of $R_{yx}^{2\omega}$ in the superconducting region and the magnitude difference of nonlinear transport between the superconducting and normal states (see Supplementary Material, Section IV).

In the Supplementary Materials Section VI, we also discuss the nonreciprocal transport under a magnetic field[7–14] to obtain a comprehensive understanding of the vortex dynamics in this material (Supplementary Fig. S6). The directional dependence of the antisymmetric second-order nonlinear magnetoresistance is rotated by 90° from the case under time-reversal symmetry, which further supports the intrinsic nature of the signals. Significantly, the theoretical estimation of the magnitude of the nonlinear magnetotransport is consistent with the experimental results, as explained in Section IV in Supplementary Material. This result also supports the above scenario, based on the asymmetric vortex Hall effect.

In Fig. 4a, we compare the nonlinear anomalous Hall effects in the normal and SC states. The temperature dependence of $\frac{|E_y^{(2)}|}{\left(E_x^{(1)}\right)^2}$ in both the normal state (blue; $I$ = 4.3 mA) and the SC state (red; $I$ = 100 μA) (left) are plotted as well as $R_{xx}^\omega$ at $I$ = 100 μA (right). Note that superconductivity is destroyed even below $T_c$ when a large current ($I$ = 4.3 mA) is applied. The obtained values of $\frac{|E_y^{(2)}|}{\left(E_x^{(1)}\right)^2}$ below $T_c$ are smoothly connected to the normal state contribution. $\frac{|E_y^{(2)}|}{\left(E_x^{(1)}\right)^2}$ in the SC state at $I$ = 100 μA indicates a remarkable enhancement by orders of magnitude compared to that in the normal state. A similar gigantic enhancement of second-order nonlinear transport is also observed in the nonreciprocal magnetotransport[10–12,14], implying that nonlinear transport is universally enhanced in the SC state, regardless of the time-reversal symmetry being preserved or not.

Finally, we compare the nonlinear anomalous Hall effect observed in the present system of PbTaSe$_2$ and with those previously reported for few-layer WTe$_2$[19,20] and TaIrTe$_4$[21]. In Fig. 4b, the values of $\frac{|E_y^{(2)}|}{\left(E_x^{(1)}\right)^2}$ are plotted as a function of $(\sigma_{xx}^\omega)^2$ for all materials. Similar plots of anomalous transverse signal versus longitudinal conductivity are known to be useful for discussing the mechanisms of the linear anomalous Hall effect in itinerant magnets[41] and anomalous thermal Hall effect in insulators[42]. In bilayer (2L) WTe$_2$, the conductivity is small because the Fermi level is located near the band edge and the behavior of the nonlinear anomalous Hall effect is rather complex, even exhibiting a sign change depending on the Fermi level position and the electrical displacement field. The observed nonlinear anomalous Hall effect can be explained well by the Berry curvature dipole effect in this case[19]. In the few-layer WTe$_2$ or TaIrTe$_4$, conductivity increases and both the skew scattering mechanism and the Berry curvature dipole effect are discussed as the origin of the nonlinear anomalous Hall effect, which has already been discussed in the previous paragraph. The magnitude of $\frac{|E_y^{(2)}|}{\left(E_x^{(1)}\right)^2}$ in the few-layer WTe$_2$ and TaIrTe$_4$ was approximately $10^{-3} \sim 10^{-2}$ μmV$^{-1}$. In our PbTaSe$_2$ samples, the conductivity is significantly larger than that in WTe$_2$, and $\frac{|E_y^{(2)}|}{\left(E_x^{(1)}\right)^2}$ also shows large values even in the normal state ($\frac{|E_y^{(2)}|}{\left(E_x^{(1)}\right)^2}$ approximately $10^2$ μmV$^{-1}$). Interestingly, it appeared that the data of the few-layer WTe$_2$, bulk TaIrTe$_4$, and present PbTaSe$_2$ were aligned in one line in this plot, potentially revealing the universal feature of the scattering-induced nonlinear anomalous Hall effect. Moreover, the $\frac{|E_y^{(2)}|}{\left(E_x^{(1)}\right)^2}$ values became even larger by two orders of magnitude in the SC state. Although we cannot simply compare the nonlinear anomalous Hall effect in the normal state and that in the SC state, we can clearly acknowledge the remarkable enhancement of the nonlinear anomalous transport in the SC region in Figs. 4a and 4b. These results imply that the large conductivity in the normal state and the vortex dynamics in the SC state may be advantageous for giant anomalous nonlinear transport.

In summary, we studied the second-order nonlinear transport in trigonal superconductor PbTaSe$_2$ under the time-reversal symmetric condition. The observed nonlinear anomalous Hall effect and intrinsic rectification effect satisfy the characteristic directional dependence of the trigonal symmetry. Furthermore, both signals are significantly enhanced around the superconducting transition, where the excitation of vortex/antivortex string pairs governs the resistance. The asymmetric vortex Hall effect is responsible for the observed nonlinear transport. The present results elucidate a new aspect of vortex dynamics in superconductors and pave the way for investigating new properties and functionalities in noncentrosymmetric conductors.

**Methods**

**Device fabrication**

Bulk PbTaSe$_2$ single crystals were grown using a flux method in an evacuated quartz tube. Stoichiometric amounts of Pb, Ta, and Se were sealed in an evacuated quartz tube, and 50 mol% KCl and 50 mol% PbCl$_2$ were mixed. The quartz tube was heated at 900 °C for 24 h and then cooled to room temperature. After crystal growth, the flux was removed by dissolution in water. The obtained PbTaSe$_2$ single crystals were exfoliated into thin flakes using the Scotch-tape method, and the flakes were transferred onto a Si/SiO$_2$ substrate. The thickness of the exfoliated flakes was measured using atomic force microscopy. A Hall bar configuration was fabricated on the flakes with Au (150 nm)/Ti (9 nm) electrodes. The pattern was fabricated using electron beam lithography, and the electrodes were deposited using an evaporator.

In fabricating the Hall bar configuration on the exfoliated flakes, we judged the crystal orientation from the straight edges of the flakes, which can be assumed to be in the zigzag direction. It is known that straight edges in exfoliated transition-metal dichalcogenides are identical to zigzag directions with high probability[28]. Although PbTaSe$_2$ has intercalated Pb layers in TaSe$_2$, we also adopted this criterion to determine the crystal orientation of PbTaSe$_2$. After the transport measurement of sample 1, it was double checked by the STEM measurement, as discussed in the main text. From the results of the STEM measurement for sample 1, we conclude that the above method of determining the crystal orientation can also be applied to PbTaSe$_2$. Schematic images of PbTaSe$_2$ in the main text are drawn by VESTA[43].

**Transport measurements**

The first and second harmonic resistances were measured using AC lock-in amplifiers (Stanford Research Systems Model SR830 DSP) with a frequency of 13 Hz in a quantum design physical property measurement system.

As discussed in previous studies[1–14], the voltage in the noncentrosymmetric system can be given as follows:

$$V = R^{(1)}I + R^{(2)}I^2,$$

where the first and second terms represent linear and second-order nonlinear transport, respectively. In this study, we focus mainly on $R^{(2)}$ under a time-reversal symmetric condition, that is, without a magnetic field.

When an AC bias current with a frequency of $\omega$ ($I = I_0 \sin \omega t$) is applied, it leads to

$$V = R^{(1)} I_0 \sin \omega t + R^{(2)} I_0^2 \sin^2 \omega t$$

$$= R^{(1)} I_0 \sin \omega t + \frac{1}{2} R^{(2)} I_0^2 \left\{ 1 + \sin\left(2\omega t - \frac{\pi}{2}\right) \right\}$$

Therefore, by extracting the first and second harmonic resistances, we obtain

$$R^\omega \equiv \frac{V^\omega}{I_0} = R^{(1)}$$

and

$$R^{2\omega} \equiv \frac{V^{2\omega}}{I_0} = \frac{1}{2} R^{(2)} I_0.$$

Next, we derive the expression for the normalized nonlinear anomalous Hall effect $\frac{E_y^{(2)}}{\left(E_x^{(1)}\right)^2}$, where $E_y^{(2)}$ and $E_x^{(1)}$ are the second-order nonlinear electric fields in the transverse direction and the linear electric field in the longitudinal direction, respectively, when current is applied along the zigzag direction. $E_y^{(2)}$ is written as

$$E_y^{(2)} = \rho_{yx}^{(2)} j_x^2,$$

where $j_x$ is the current density and $\rho_{yx}^{(2)}$ is the second-order resistivity. By considering $V_y^{(2)} = W E_y^{(2)}$ and $I_x = W t j_x$, where $V_y^{(2)}$ is the nonlinear Hall voltage, $W$ is the channel width, $t$ is the thickness of the flake, and $I_x$ is the current, it transforms into

$$V_y^{(2)} = \frac{\rho_{yx}^{(2)}}{Wt^2} I_x^2 = R_{yx}^{(2)} I_x^2$$

Therefore, using $E_x^{(1)} = \rho_{xx}^{(1)} j_x$ , where $\rho_{xx}^{(1)} = \frac{Wt}{L} R_{xx}^{(1)}$ is the linear longitudinal resistivity with channel length $L$, $\frac{E_y^{(2)}}{\left(E_x^{(1)}\right)^2}$ is calculated as

$$\frac{E_y^{(2)}}{\left(E_x^{(1)}\right)^2} = \frac{\rho_{yx}^{(2)}}{\left(\rho_{xx}^{(1)}\right)^2} = \frac{L^2}{W} \frac{R_{yx}^{(2)}}{\left(R_{xx}^{(1)}\right)^2} = \frac{2L^2}{WI_0} \frac{R_{yx}^{2\omega}}{(R_{xx}^{\omega})^2}.$$

**Selection rules for nonlinear transport under time-reversal symmetric condition in trigonal systems**

Nonlinear current density $\boldsymbol{j}^{(2)}$ in the noncentrosymmetric system is generally written as $\boldsymbol{j}^{(2)} = \boldsymbol{\beta EE}$ or $j_i^{(2)} = \beta_{ijk} E_j E_k$, where $\beta$ is a third-order tensor[44]. Considering PbTaSe$_2$ with point group D$_{3h}$, $\beta$ leads to

$$\beta = \begin{pmatrix} \beta_{11} & -\beta_{11} & 0 & 0 & 0 & 0 \\ 0 & 0 & 0 & 0 & 0 & -\beta_{11} \\ 0 & 0 & 0 & 0 & 0 & 0 \end{pmatrix}$$

Here, $x$, $y$, and $z$ are parallel to the armchair direction, parallel to the zigzag direction, and perpendicular to the plane, respectively. Therefore, $\boldsymbol{j}^{(2)}$ under electric field $\boldsymbol{E}$ is written as follows:

$$\boldsymbol{j}^{(2)} = \begin{pmatrix} \beta_{11}(E_x^2 - E_y^2) \\ -2\beta_{11} E_x E_y \\ 0 \end{pmatrix}$$

When the electric field is applied along the armchair direction ($E_x = E, E_y = 0$), $\boldsymbol{j}^{(2)}$ leads to

$$\boldsymbol{j}^{(2)} = \begin{pmatrix} \beta_{11} E^2 \\ 0 \\ 0 \end{pmatrix}$$

On the other hand, for the electric field applied along the zigzag direction ($E_x = 0, E_y = E$), $\boldsymbol{j}^{(2)}$ leads to

$$\boldsymbol{j}^{(2)} = \begin{pmatrix} -\beta_{11}E^2 \\ 0 \\ 0 \end{pmatrix}$$

In both cases, $\boldsymbol{j}^{(2)}$ has only the *x*-component. This directional dependence (selection rule) in trigonal systems was confirmed in this study.

**Acknowledgments**

We thank T. Nojima, S. Koshikawa, H. Isobe, and N. Nagaosa for fruitful discussions. Y.M.I. was supported by the Advanced Leading Graduate Course for Photon Science (ALPS). T.I. was supported by JSPS KAKENHI grant numbers JP19K21843, JP19H01819, JP20H05264, grant from Yazaki Memorial Foundation for Science and Technology, JST PRESTO (grant no. JPMJPR19L1). T.S. was supported by JST CREST (grant no. JPMJCR16F2). This work was supported by JSPS KAKENHI grant number JP19H05602 and the A3 Foresight Program.

**Author contributions:** Y.M.I., T.I., and Y.I. conceived the research project. H.N. and T.S. synthesized the bulk material. Y.M.I. and C.G. fabricated the microdevices, performed the experiments, and analyzed the data. S.H. performed the theoretical calculations. Y.M.I., T.I., S.H., and Y.I. wrote the manuscript. All authors have led the physical discussions.

**Competing financial interests:** The authors declare that they have no competing interests.

**Data and materials availability:** All data needed to evaluate the conclusions of this study are available as Supplementary Material.

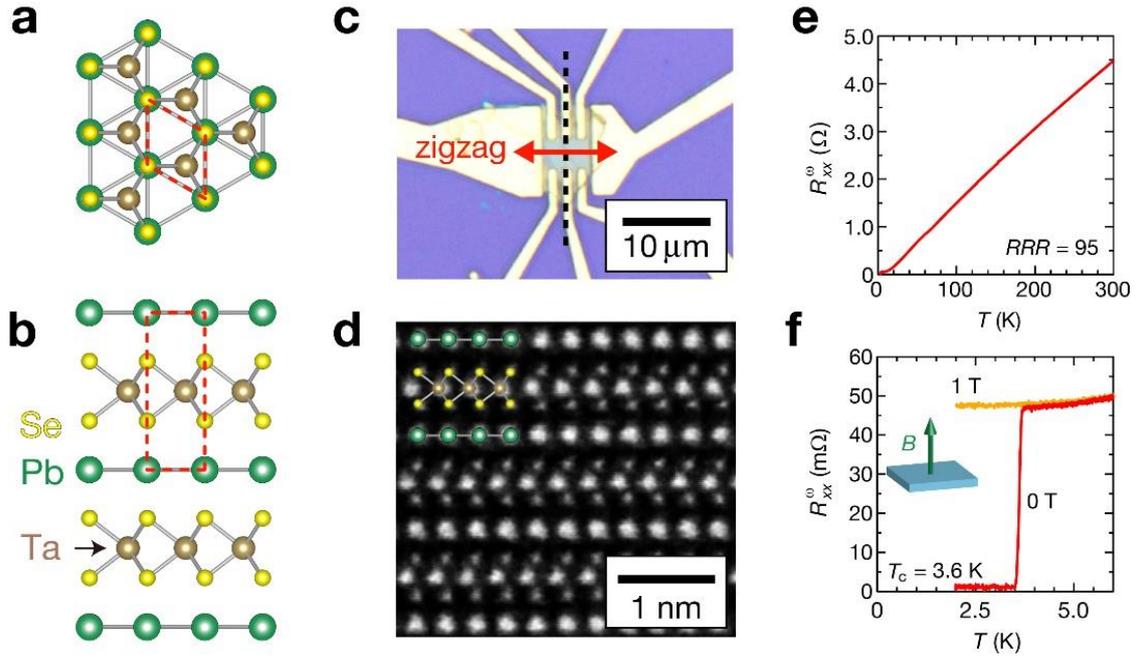

**Figure 1. Crystal structure and superconducting properties of PbTaSe$_2$.**

**a,b**, (**a**) Top and (**b**) side views of PbTaSe$_2$. Pb layers are intercalated in TaSe$_2$ with 1H stacking. Red dashed squares indicate the unit cell. **c**, Optical microscope image of the PbTaSe$_2$ device of sample 1. **d**, A cross-sectional scanning transmission electron microscope (STEM) image of PbTaSe$_2$ along the dashed line in Fig. 1c. Schematic of the cross section of PbTaSe$_2$ along the armchair direction is also displayed. **e**, Temperature dependence of the first harmonic resistance $R_{xx}^\omega$ in sample 2. **f**, Temperature dependence of $R_{xx}^\omega$ around the superconducting transition ($T_c$ = 3.6 K) in sample 2 when $I$ = 140 μA. Red and orange lines depict the data under 0 T (red) and 1 T (orange), respectively. Magnetic field $B$ is applied perpendicular to the 2D layers.

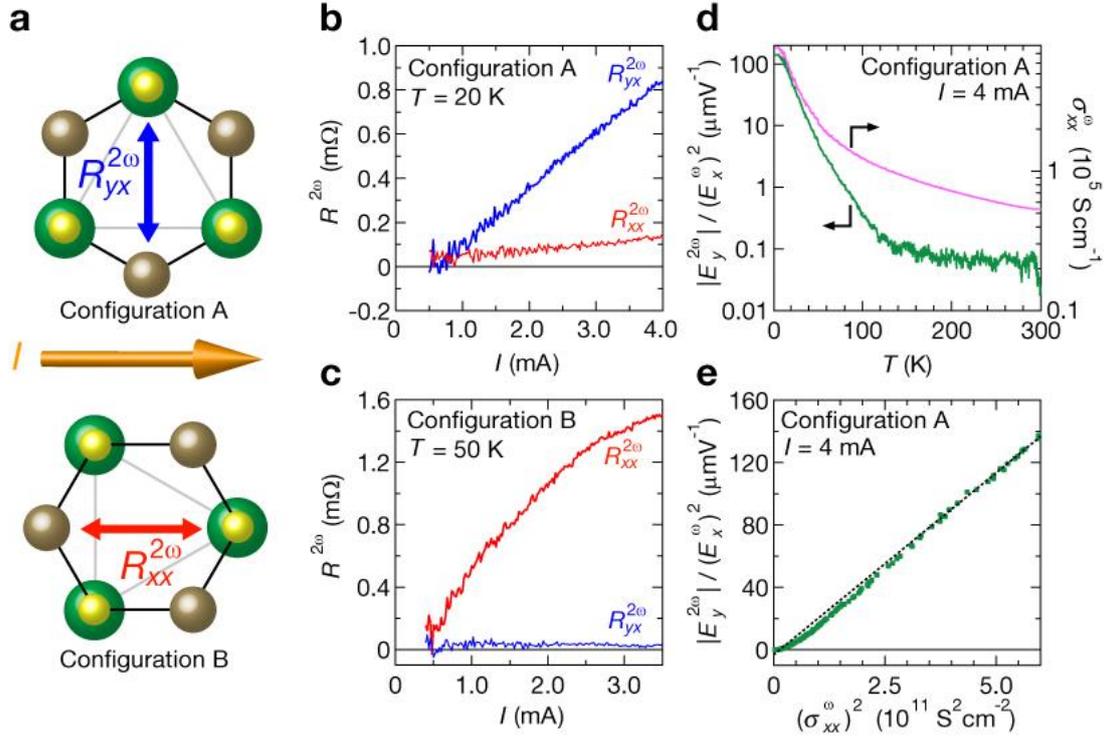

**Figure 2. Current and temperature dependences of second-order anomalous transport in normal state. a**, Directional dependence (selection rule) of the nonlinear Hall/rectification effect in trigonal PbTaSe$_2$. In configuration A (B), where the current is applied along the zigzag direction (armchair direction), the second harmonic signal is expected in the transverse (longitudinal) direction. **b,c**, Current dependences of the second harmonic resistance $R^{2\omega}$ in (**b**) configurations A (sample 3) at $T = 20$ K and (**c**) configuration B (sample 4) at $T = 50$ K. Red and blue lines indicate longitudinal ($R^{2\omega}_{xx}$) and transverse ($R^{2\omega}_{yx}$) resistance, respectively. Directional dependence of the nonlinear transport illustrated in Fig. 1(a) is confirmed. **d**, Normalized second harmonic response $\frac{|E_y^{(2)}|}{\left(E_x^{(1)}\right)^2}$ (green, left) and first harmonic conductivity $\sigma^{\omega}_{xx}$ (pink, right) as a function of temperature at $I = 4$ mA in configuration A (sample 1). **e**, Normalized second harmonic resistivity $\frac{|E_y^{(2)}|}{\left(E_x^{(1)}\right)^2}$ as a function of $(\sigma^{\omega}_{xx})^2$ in configuration A (sample 1). Black dotted line indicates the linear fitting in the low-temperature (high-conductivity) region.

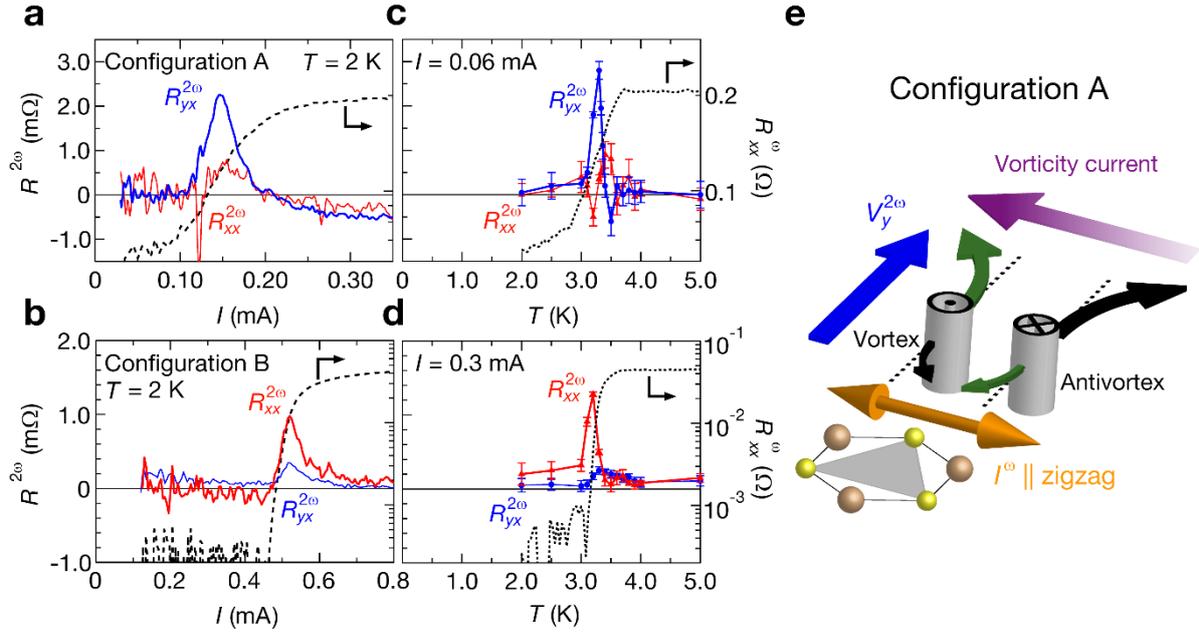

**Figure 3. Current and temperature dependences of first/second harmonic signals around superconducting transition and schematic of asymmetric vortex Hall effect as a possible origin for nonlinear transport. a,b**, Current dependence of $R^{2\omega}$ (left) and $R_{xx}^{\omega}$ (right) at $T = 2$ K in (**a**) configuration A (sample 1) and (**b**) configuration B (sample 2). Red and blue lines indicate longitudinal ($R_{xx}^{2\omega}$) and transverse ($R_{yx}^{2\omega}$) resistance, respectively. **c,d**, Temperature dependences of $R^{2\omega}$ (left) and $R_{xx}^{\omega}$ (right) in (**c**) configuration A (sample 1) and (**d**) configuration B (sample 2) directions. The current value is 0.06 mA and 0.3 mA in Figs. 3c and 3d, respectively. Red and blue lines indicate longitudinal ($R_{xx}^{2\omega}$) and transverse ($R_{yx}^{2\omega}$) resistances, respectively. **e**, Schematic of the rectified vortex/antivortex Hall effect in configuration A. Black (green) arrows denote the trajectories of the vortices/antivortices Hall effect when current flows along the zigzag direction. The rectification of vortices/antivortices reflecting the trigonal potential is represented by the thickness differences of arrows. Purple arrow indicates the antiparallel motions of vortices or vorticity current. Nonlinear voltage ($V_y^{2\omega}$) appears perpendicular to the vorticity current, in analogy to the inverse spin Hall effect. Regardless of the direction of the current, excess voltage with the same sign appears along the armchair direction, which can be observed as the second harmonic resistance.

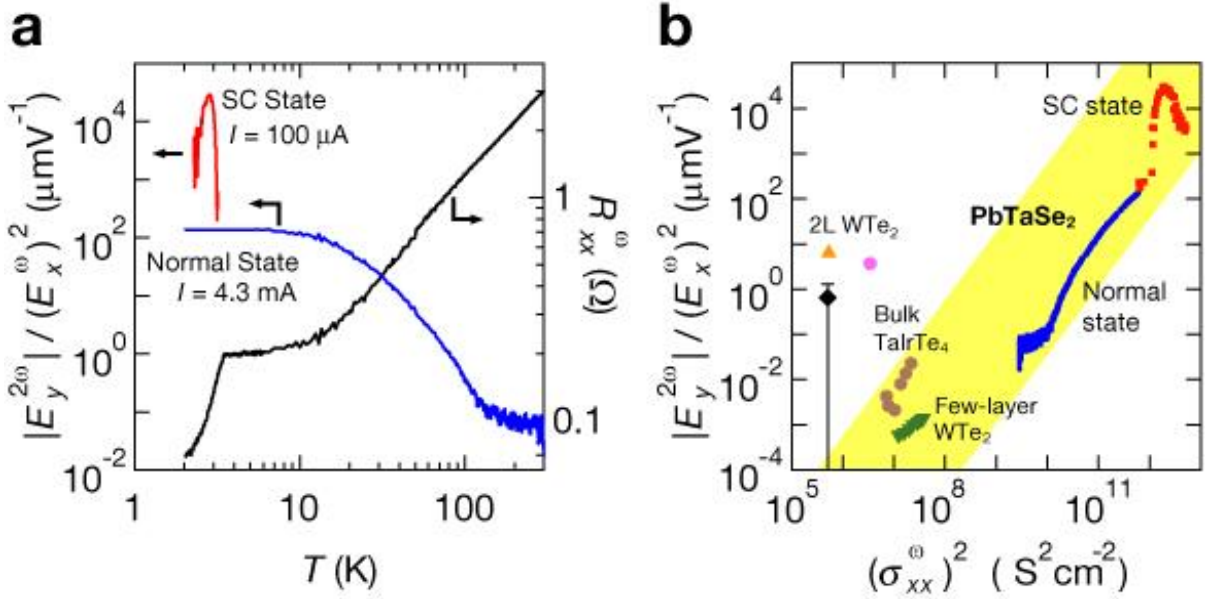

**Figure 4. Summarized temperature dependence of nonlinear Hall effect in PbTaSe$_2$ and comparison of nonlinear anomalous Hall effects in few-layer WTe$_2$ and TaIrTe$_4$. a**, Temperature dependences of $\frac{|E_y^{(2)}|}{(E_x^{(1)})^2}$ in the normal state (blue; $I$ = 4.3 mA) and in the superconducting (SC) state (red; $I$ = 100 μA) (left), and $R_{xx}^{\omega}$ at $I$ = 100 μA (right) in sample 1. **b**, $\frac{|E_y^{(2)}|}{(E_x^{(1)})^2}$ as a function of temperature in PbTaSe$_2$, few-layer WTe$_2$[19,20], and TaIrTe$_4$[21]. Blue and red squares indicate $\frac{|E_y^{(2)}|}{(E_x^{(1)})^2}$ of PbTaSe$_2$ (sample 1) in the normal state and the SC state, respectively. Green triangles depict $\frac{|E_y^{(2)}|}{(E_x^{(1)})^2}$ in the few-layer WTe$_2$. Data were sourced from Kang et al.[20]. Black diamond, pink circle, and orange triangle represent $\frac{|E_y^{(2)}|}{(E_x^{(1)})^2}$ in the bilayer (2L) WTe$_2$. These were calculated using data from Ma et al.[19]. Carrier densities *n* are approximately 0 cm$^{-2}$ (black), -7 × 10$^{12}$ cm$^{-2}$ (pink), and -7 × 10$^{11}$ cm$^{-2}$ (orange) (*n* values corresponding to carrier densities where the nonlinear anomalous Hall effect exhibits the local maximum values). Brown circles depict $\frac{|E_y^{(2)}|}{(E_x^{(1)})^2}$ in the 16 nm-thick TaIrTe$_4$. Data were sourced

from Kumar et al.[21].

# Supplementary Materials for

# Giant intrinsic rectification and nonlinear Hall effect under time−reversal symmetry in a trigonal superconductor

## I. Linear magnetotransport of PbTaSe₂

In Figs. S1a and S1b, we show the magnetic field dependence of longitudinal resistivity $\rho_{xx}$ and transverse resistivity $\rho_{yx}$, respectively, at $T = 8$ K in sample 5. $\rho_{xx}$ and $\rho_{yx}$ are calculated by $\rho_{xx} = \frac{Wt}{L} R_{xx}^{sym}$ and $\rho_{yx} = tR_{yx}^{asym}$, respectively, where $R_{xx}^{sym}/R_{yx}^{asym}$ the symmetrized longitudinal resistance/antisymmetrized transverse resistance as a function of the magnetic field, $W = 1.8$ μA the width, $L = 2.7$ μA the length and $t = 63$ nm the thickness for sample 5. $\rho_{xx}(B)$ shows the simple positive magnetoresistance while $\rho_{yx}(B)$ shows the complex multi-carrier behavior. At the high field region, positive signal is dominant in $\rho_{yx}(B)$.

We analyzed $\rho_{yx}(B)$ by using the two-carrier model[1,2] (yellow dashed line in Fig. S1b)

$$\rho_{yx}(B) = \frac{B}{e} \frac{(n_h \mu_h^2 - n_e \mu_e^2) + \mu_h^2 \mu_e^2 (n_h - n_e) B^2}{(n_h \mu_h + n_e \mu_e)^2 + \mu_h^2 \mu_e^2 (n_h - n_e)^2 B^2} \tag{S1}$$

where $n_h$ ($n_e$) and $\mu_h$ ($\mu_e$) are the carrier density and mobility of holes (electrons), respectively. Obtained fitting parameters are $n_h = 5.7 \times 10^{22}$ cm⁻³, $n_e = 4.6 \times 10^{22}$ cm⁻³, $\mu_h = 2.1 \times 10^4$ cm²V⁻¹s⁻¹ and $\mu_e = 2.2 \times 10^4$ cm²V⁻¹s⁻¹. Carrier density of hole is in the same order of one in 2H-TaSe₂ ($3.5 \times 10^{22}$ cm⁻³)[3], which implies that holes are mainly originating from TaSe₂ layers.

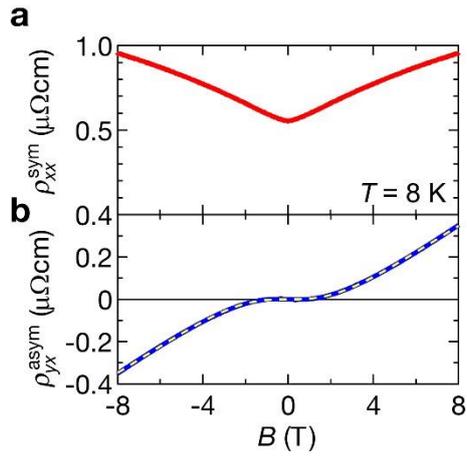

**Figure S1. Linear magnetotransport of PbTaSe$_2$. (a, b)** Longitudinal resistivity $\rho_{xx}$ (a) and transverse resistivity $\rho_{yx}$ (b) at $T$ = 8 K in sample 5. Yellow dashed line shows the fitting curve for $\rho_{yx}(B)$ by the two-carrier model.

## II. Vortex Hall effect probed by linear Hall effect in PbTaSe$_2$

In the layered superconductors, the difference in chemical potentials in the normal core and the other superconducting region leads to the charging of vortices, which causes transverse motion of vortices and resultant anomalous sign reversal in Hall resistance near the superconducting transition[4–10]. Figures S2a and S2b show the temperature dependence of symmetrized $R_{xx}^{\omega}$ ($R_{xx}^{\text{sym}}$, Fig. S2a), antisymmetrized $R_{xx}^{\omega}$ ($R_{xx}^{\text{asym}}$, red) and $R_{yx}^{\omega}$ ($R_{yx}^{\text{asym}}$, blue) (Fig. S2b) at $B = \pm 0.003$ T and $I = 10$ μA, respectively, for sample 5. During the superconducting transition, $R_{yx}^{\text{asym}}$ shows the large negative peak. The sign of the Hall effect is opposite to $R_{yx}^{\text{asym}}$ in the normal state, which is much smaller than the negative component as seen in Fig. S1b, because we applied the magnetic field of only 0.003 T in order not to suppress superconductivity. Such a peak is absent in $R_{xx}^{\text{asym}}$. The sign reversal in the Hall resistance between the normal state and the superconducting state strongly suggests the occurrence of the vortex Hall effect in this layered superconductor. Figures S2c and S2d display the magnetic field dependence of $R_{xx}^{\text{sym}}$ and $R_{yx}^{\text{asym}}$, respectively, at $T = 3.1$ K (purple) and 3.4 K (orange) for sample 6. The current value was set at $I = 50$ μA in this measurement. The negative peak is clearly observed around the superconducting transition for both cases, which are consistent with the above scenario of the vortex Hall effect. We also calculated the Hall angle of vortex Hall effect as $r = \frac{R_{yx}^{\text{asym}}(B)}{R_{xx}^{\text{sym}}(B)} \frac{L}{W} = 0.005\text{-}0.01$ (see Supplementary material section IV in detail), which is consistent with the other layered superconductors. Thus, we conclude that the Hall anomaly observed in PbTaSe$_2$ is attributed to the vortex Hall effect.

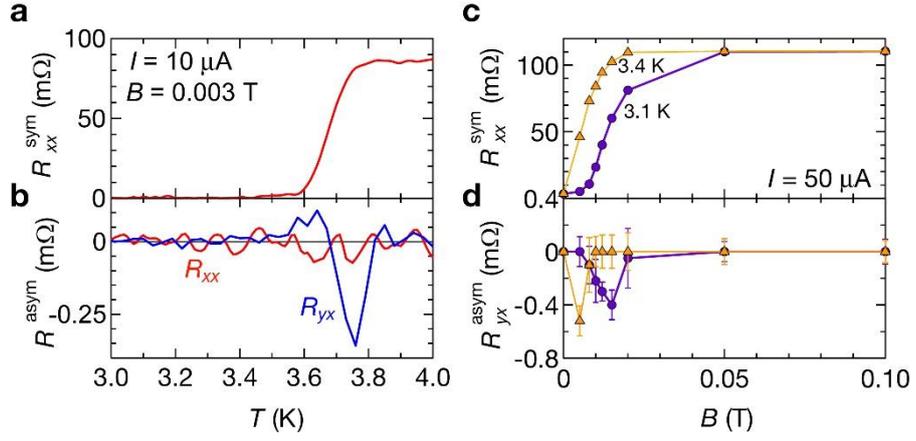

**Figure S2. Vortex Hall effect probed by linear Hall effect in PbTaSe$_2$.** (**a**, **b**), Temperature dependence of $R_{xx}^{\text{sym}}$ (a) and $R^{\text{asym}}$ (b) at $I$ = 10 μA and $B$ = ±0.003 T in sample 5. Red (blue) curve in Fig. S2b shows $R_{xx}^{\text{asym}}$ ($R_{yx}^{\text{asym}}$). (**c**, **d**) Magnetic field dependence of $R_{xx}^{\text{sym}}$ (c) and $R_{yx}^{\text{asym}}$ (d) at $I$ = 50 μA and $T$ = 3.1 K (purple), 3.4 K (orange) in sample 6.

**III. Asymmetric vortex Hall effect as the origin of the nonlinear transport**

In Figs. S3a and S3b, we propose a possible mechanism of the observed nonlinear anomalous Hall effect (Fig. S3a) and rectification effect (Fig. S3b) in trigonal superconductors by considering the vortex/antivortex Hall effect on the trigonal potential. In the present samples of layered superconductor PbTaSe$_2$ of typically 80 nm in thickness, vortex/antivortex pairs are excited in the form of vortex string (Fig. S3e) by temperature or current, as we explained later. We assumed that these excited vortex/antivortex pairs are subjected to the vortex Hall effect even in the absence of magnetic fields, where they are driven by their own magnetic flux in the same manner as in the presence of external magnetic field, discussed in the previous section (Supplementary Materials II). When the current is applied along the zigzag (armchair) direction, which corresponds to configuration A (B), vortices/antivortices are first driven to the armchair (zigzag) direction and then curved in the transverse zigzag (armchair) direction due to the vortex/antivortex Hall effect. The black/green arrows denote the motions of vortices/antivortices when the current flows leftward (rightward). During these processes, vortices/antivortices flows are rectified by the trigonal potential, which are drawn by the thickness differences of black and green arrows in Figs. S3a and S3b. Thus, in configuration A (B), vortices/antivortices flow in parallel along the armchair (zigzag) direction and in antiparallel along the zigzag (armchair) direction. Parallel motion of vortices and antivortices cancel with each other, whereas antiparallel motions of vortices/antivortices leads to the "vorticity current", which can be detected as the dc voltage in the direction perpendicular to those motions, causing the finite $R_{yx}^{2\omega}$ ($R_{xx}^{2\omega}$), or nonlinear Hall effect (rectification effect) in configuration A (B). We call this antiparallel motion "vorticity current" in analogy to the spin current in the inverse spin Hall effect. It is noted that the effect of vortices and antivortices driven in the same direction are canceled out.

In above consideration, we assumed the existence of free vortex-antivortex in PbTaSe$_2$

in the absence of external magnetic field. This assumption is supported by the following model calculation based on Ref. 12. In the layered superconductors, three types of excitations are known to occur[11,12]: (i) vortex-antivortex pair (Fig. S3c), (ii) vortex ring (Fig. S3d) and (iii) vortex string pair (Fig. S3e). Vortex-antivortex pairs are vortex loops excited in each layer, which are independent from other ones in adjacent layers (Fig. S3c). Vortex rings are vortex loops penetrating multiple layers within a sample (Fig. S3d). Vortex string pairs are vortex loops penetrating all the layers (Fig. S3e). Among these kinds of excitations, vortex string pairs are considered to be the origin of BKT-like transition in the layered superconductor $YBa_2Cu_3O_7$, in 10 layer-thick films[12]. Because the anisotropy parameter $\gamma = B_{c2}^{\parallel}/B_{c2}^{\perp}$ ($B_{c2}^{\parallel}$ and $B_{c2}^{\perp}$ are in-plane and out-of-plane critical magnetic field, respectively) of $YBa_2Cu_3O_7$ ($\gamma \lesssim 10$)[12] is in the same order as the one of $PbTaSe_2$ ($\gamma \sim 5$)[13], similar vortex excitations might exist at zero magnetic field in the present $PbTaSe_2$.

To evaluate the existence of the vortex string pairs in $PbTaSe_2$, we estimated the energies of vortex-antivortex pair ($U_{vp}$), vortex ring ($U_{ring}$) and vortex string pair ($U_{str}$). According to the previous studies[11,12], these energies are described as

$$U_{vp} = K \ln\left(\frac{r_v}{\xi_{ab}}\right) + K_\perp \left(\frac{r_v}{d}\right)^2 + 2E_c \quad (S2)$$

$$U_{ring} = \frac{r_v}{d}\left[K \ln\left(\frac{r_v}{\xi_{ab}}\right) + 2E_c\right] \quad (S3)$$

$$U_{str} = n\left[K \ln\left(\frac{r_v}{\xi_{ab}}\right) + 2E_c\right] \quad (S4)$$

where $r_v$ is the diameter of vortex loop, $K$ and $K_\perp$ are the intralayer and interlayer coupling constant, respectively, $\xi_{ab}$ is the in-plane coherence length, $d$ is the thickness of one layer, $n$ is the number of layers and $E_c$ is the energy to create one vortex in one layer. $K$ and $K_\perp$ is related to anisotropy parameter $\gamma$ as $K/K_\perp = \gamma^2$ and we ignore $E_c$ because it is small. Figures S3f-h show $U_{vp}$ (purple), $U_{ring}$ (green) and $U_{str}$ (blue) divided by $K$, as a function of $r_v/\xi_{ab}$ at several temperatures. Here, we used $K/K_\perp = \gamma^2 = 25$, $\xi_{ab} = \xi_{ab0}\sqrt{\frac{T_c}{T_c-T}}$, where $\xi_{ab0} = 8$ nm and $T_c = 3.8$ K, $d = 1$ nm and $n = 100$ in the present case of PbTaSe$_2$[13]. At $T/T_c = 0$ (Fig. S3f), vortex ring has the lowest energy at small $r_v$ region. However, at higher temperatures of $T/T_c = 0.5$ and 0.7 (Figs. S3g and S3h), $U_{str}$ decreases and becomes lower than $U_{ring}$, indicating that the vortex string pair is the lowest energy excitation in PbTaSe$_2$ above $T_c/2$. This result is explained in terms of the enhancement of coherence length, that is, the vortex radius toward $T_c$, which makes the system more 2D-like. Thus, we conclude that vortex string pairs are excited in PbTaSe$_2$ flake near the superconducting transition and serve as free vortices and antivortices at zero magnetic field to cause the nonlinear superconducting transport.

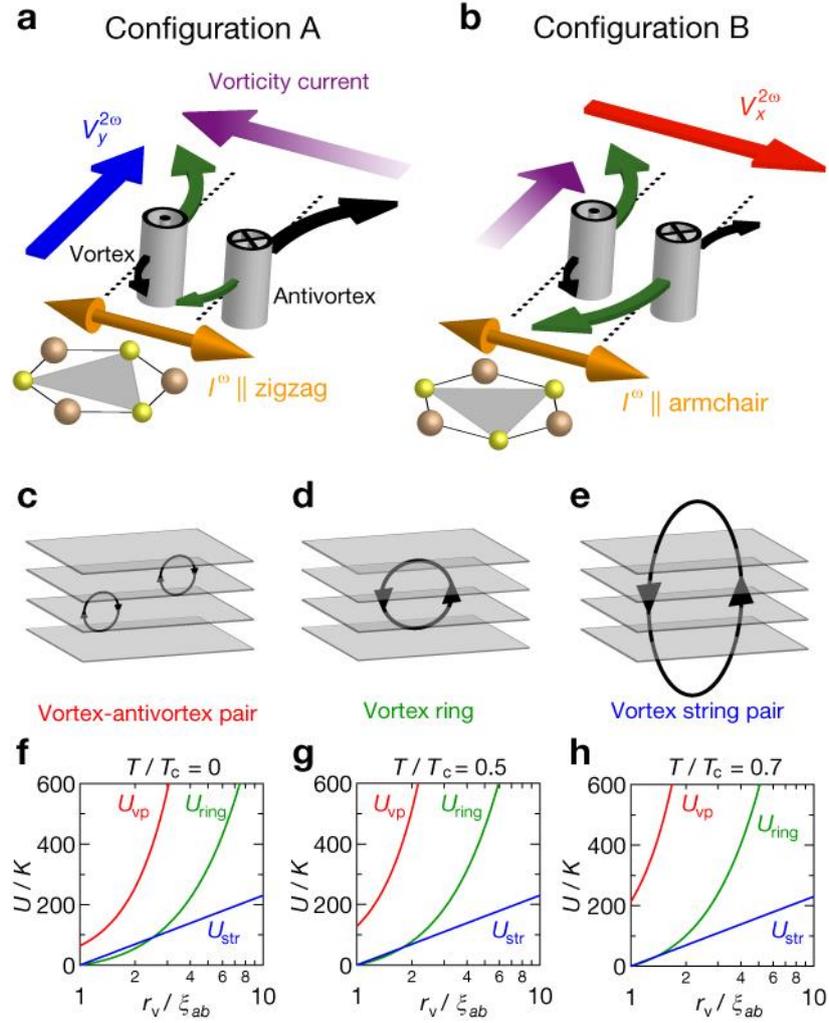

**Figure S3. Schematic images of asymmetric vortex Hall effect and vortex string pairs in PbTaSe$_2$.**

(**a, b**), Schematic images of rectified charged vortices/antivortices when the current is applied along zigzag (a) and armchair (b) directions. Black (green) arrows denote the trajectories of charged vortices/antivortices when the current flows leftward (rightward). The rectification of vortices/antivortices reflecting the trigonal potential is drawn by the thickness differences of arrows in the figures. Purple arrows show the antiparallel motions of vortices, or vorticity current. Nonlinear signals appear perpendicular to the vortex spin current, in analogy to the inverse spin Hall effect. In configuration A (B), finite nonlinear transverse voltage $V_y^{2\omega}$ (nonlinear longitudinal voltage $V_x^{2\omega}$) is observed. (**c-e**), Schematic images of vortex-antivortex pair (c), vortex ring (d) and vortex string pair (e) in the layered superconductor. (**f-h**), Excitation energy of vortex-antivortex pair ($U_{vp}$), vortex ring ($U_{ring}$) and vortex string pair ($U_{str}$) divided by intralayer coupling constant $K$ at $T/T_c = 0$ (f), 0.5 (g) and 0.7 (h).

**IV. Theoretical description of vortex-induced linear and nonlinear transport**

We describe the transport coefficients in terms of vortex contributions. With the weak force $\boldsymbol{F} = (F_x, F_y)$ acting on a vortex, the velocity for the vortex (vorticity $s = +$) or antivortex ($s = -$) is given by

$$\boldsymbol{v}^s = q_1 \begin{pmatrix} F_x \\ sF_y \end{pmatrix} + q_2 \begin{pmatrix} 2sF_x F_y \\ F_x^2 - F_y^2 \end{pmatrix} \quad (S5)$$

The second term in the right-hand side originates from the trigonal symmetry with the configuration A. The motions of vortex and antivortex are switched by the mirror transformation with respect to the $x$ axis. The voltage differences along $x$ and $y$ directions are given by the Josephson relation[14]

$$V_x = \phi_0^* L \sum_{s=\pm} s n^s v_y^s, \quad V_y = \phi_0^* W \sum_{s=\pm} s n^s v_x^s \quad (S6)$$

where $\phi_0^* = h/2|e|$ is the superconducting flux quantum, and $n^s$ is the number density of vortices or antivortices. The sample lengths along $x$ and $y$ directions are $L$ and $W$, respectively. Thus, the voltage is proportional to the vorticity current.

Next, we determine the concrete form of the force $\boldsymbol{F}$. For the linear response, the driving force $j\phi_0^*$ on a vortex, where $j$ is the current density, is balanced by the force from environment as

$$sj\phi_0^* \hat{\boldsymbol{y}} = \eta_\parallel \boldsymbol{v}^s + s\eta_\perp \hat{\boldsymbol{z}} \times \boldsymbol{v}^s \quad (S7)$$

where $\eta_\parallel$ is the friction coefficient and $\eta_\perp$ is responsible for the vortex Hall effect. The hat symbol represents a unit vector. Assuming that the Hall effect is small, i.e., $r \equiv \frac{\eta_\perp}{\eta_\parallel} \ll 1$, we obtain the relations $q_1 = \frac{1}{\eta_\parallel}$, $F_y = j\phi_0^*$, and $F_x = rF_y$.

As for the ratchet effect, we employ the information from the Brownian motion of point particle in the one-dimensional asymmetric potential for simplicity[15]. We take the potential form shown in Fig. S4a, where $\ell_v$ is the periodicity of the potential for vortices, and the length $c\ell_v$ is the size of each pinning center. The potential height is given by $U$, and we will take $c = 0.1$ in the numerical evaluation. From the Fokker-Planck equation, the response coefficients are given by

$$q_1 = \frac{1}{\eta_0} g_1(\beta U), \qquad q_2 = \frac{\beta \ell_v}{\eta_0} g_2(\beta U), \quad \text{(S8)}$$

$$g_1(x) = \frac{x^2}{D(x)}, \qquad D(x) = 2c^2(\cosh x - 1) + (1-c)^2 x^2 + 2(1-c)cx \sinh x,$$

$$g_2(x) = \frac{c^2 x}{D(x)^2}[4c + (2-c)x^2 - (4c - (1-c)x^2)\cosh x - (3 - 4c)x \sinh x], \quad \text{(S9)}$$

where $\beta = \frac{1}{k_B T}$ is the inverse temperature. While the actual potential shape is dependent on the system details, the coefficients $q_{1,2}$ generically have the exponential temperature dependence $e^{-\beta U}$ at low $T$ reflecting the pinning of vortices. The constant $\eta_0$ is the friction coefficient without the pinning potentials. In the Bardeen-Stephen model[16], it is given by $\eta_0 = \frac{\pi \hbar^2 \sigma_n}{2e^2 \xi_{ab}^2}$ with the normal conductivity $\sigma_n$ and the coherence length $\xi_{ab}$.

*1. Without magnetic field*

First, we consider the case without magnetic field. In this case we have the relation $n^+ = n^- = n_v/2$ where $n_v$ is the total number density of the vortices. We then obtain the linear and nonlinear transport coefficients as

$$R_{xx}^\omega = \frac{(\phi_0^*)^2 n_v L}{W \eta_0} g_1(\beta U), \qquad R_{xx}^{2\omega} = 0 \quad \text{(S10)}$$

$$R_{yx}^\omega = 0, \qquad R_{yx}^{2\omega} = \frac{(\phi_0^*)^3 n_v r \beta \ell_v I}{W \eta_0} g_2(\beta U) \quad \text{(S11)}$$

which satisfies the selection rule in the trigonal symmetry. Here we use the electrical current $I$ instead of the current density. Note that the vortex Hall effect is essential for the nonreciprocal transport signal as seen from the presence of the factor $r = \eta_\perp/\eta_\parallel$. The ratio is written by the simple quantity

$$\frac{R_{yx}^{2\omega}}{R_{xx}^{\omega}} = \frac{\phi_0^* r \ell_v I}{k_B T L} \cdot \frac{g_2(\beta U)}{g_1(\beta U)} \quad (S12)$$

which is not influenced by the number of vortices and friction coefficient. Taking the low-temperature limit, the expression is further simplified because $\frac{g_2}{g_1} = \frac{c}{2} = \text{const.}$ can be used.

Let us evaluate the magnitude of the signals. First, we estimate the length $\ell_v$ from the information under the magnetic field at low temperature: At $B = B_{\text{pin}}$, all the vortices with the number density $B/\phi_0^*$ are trapped by the pinning centers, and the mobile vortices appear for $B > B_{\text{pin}}$[15] which generate the transport signals (see Fig. S6a of the main text). Hence the periodicity of pinning potential in the low-temperature regime is estimated as $\ell_{v0} = \sqrt{\phi_0^*/B_{\text{pin}}} \simeq 3 \times 10^{-7}$ m. On the other hand, the potential height $U$ is estimated from the critical current $I_{\text{pin}}$, where the potential is well tilted by the external force and become flat at this critical current. Then we obtain the relation $U = \phi_0^* c \ell_v I_{\text{pin}}/W$. The value at low temperature is estimated by using $I_{\text{pin},0} \simeq 400$ μA (see Fig. S4b), and we get $U_0 \simeq 40$ meV. If we considered a temperature-independent potential height, the transport signal with the magnitude observed experimentally could not be reproduced due to the exponential factor $e^{-\beta U_0}$. Hence it is necessary and natural to consider the temperature dependence of $U$ which goes to zero as $T$ approaches to the mean-field critical temperature $T_c \simeq 3.8$ K. The potential height may also be estimated as the condensation energy gain at the vortex core in the presence of the normal state at the impurity site as $U \simeq p \frac{B_c^2}{2\mu_0} \pi \xi_{ab}^2 t \propto T_c - T$ where $B_c$ is the thermodynamic critical magnetic field, $p$ the fraction of pinning points[17,18], $\mu_0$ the permeability in vacuum, and $t$ the sample thickness. In addition, we assume $\ell_v \propto \sqrt{T_c - T}$ which corresponds to the fact that the size of vortices increases as $T \to T_c$ and the potential periodicity for vortices becomes effectively shorter together with the increasing coherence length $\xi_{ab}$. With these results we obtain $I_{\text{pin}} \propto \sqrt{T_c - T}$. Indeed, the experimental result supports this temperature dependence for the critical current (Fig. S4b). In this way, we can write the temperature dependent parameters as $\ell_v(T) \simeq \ell_{v0}\sqrt{\frac{T_c-T}{T_c}}$ and $U(T) \simeq U_0 \frac{T_c-T}{T_c}$. Now we can evaluate the ratio $\frac{R_{xx}^\omega}{R_{yx}^{2\omega}}$. The temperature dependence near the transition point is shown in Fig. S4c with the parameter $r = 0.01$ and the system lengths $L = 1.7$ μm, $W = 3.5$ μm. The magnitude is comparable to the experiments within the temperature range where the signal is observed.

Next, we consider the number of vortices which is necessary for the direct evaluation of $R_{yx}^{2\omega}$. We assume that the description of two-dimensional superconductors applies to the present system. Below the Kosterlitz-Thouless transition temperature $T_{\text{KT}}$, which is very close to the mean-field transition temperature $T_c$ away from dirty limit[18,19], all the vortices are paired with antivortices. On the other hand, under the current flow, the mobile vortices are generated and the number of vortices is given at low current density by[14]

$$n_v(I) = \frac{x}{2\pi \xi_{ab}^2}\left(\frac{I}{I_0}\right)^{2+\frac{x}{2}} \quad (S13)$$

with $x = \frac{4(T_{\text{KT}}-T)}{T_{\text{KT}}}$ and $I_0 = \frac{k_B T_{\text{KT}}|e|W}{\hbar \xi_{ab}}$. If the current exceeds $I_0$, all the vortices are unpaired[20]. We can estimate $I_0$ by using the Ginzburg-Landau coherence length $\xi_{ab} = \xi_{ab0}\sqrt{\frac{T_c}{T_c-T}}$ with $\xi_{ab0} \simeq 40$ nm [13], and we see that the experimentally used current $I = 100$ μA exceeds this limit at all temperatures ($I/I_0(T) \gtrsim 10$). Hence, we assume that the existing vortices are free from vortex-antivortex binding. However, it is still not easy to estimate the number of vortices which could be generated by thermal fluctuations and current noise. Furthermore, the extrapolation of Eq. S13 for the region $I > I_0$ does not work since the number of vortices much exceeds the upper bound $\sim \frac{1}{\pi \xi_{ab}^2}$ ($\sim n_v(I = I_0)$). Then we consider the number density based on this maximum value and take $n_v = \frac{\alpha'}{\pi \xi_{ab}^2}$ where $\alpha'(< 1)$ is the reduction factor from the maximum value. The resistance itself can now be evaluated. Figure S4d shows the temperature dependence of the nonlinear Hall signal $R_{yx}^{2\omega}$ for $\alpha' = 0.5$ and with the linear resistance $0.2\Omega$ in the normal state. The characteristic peak structure and the magnitude of the signal is consistent with the experimental observation. The absence of the signal in low-temperature region is not due to the vortex-antivortex binding but to the pinned vortices.

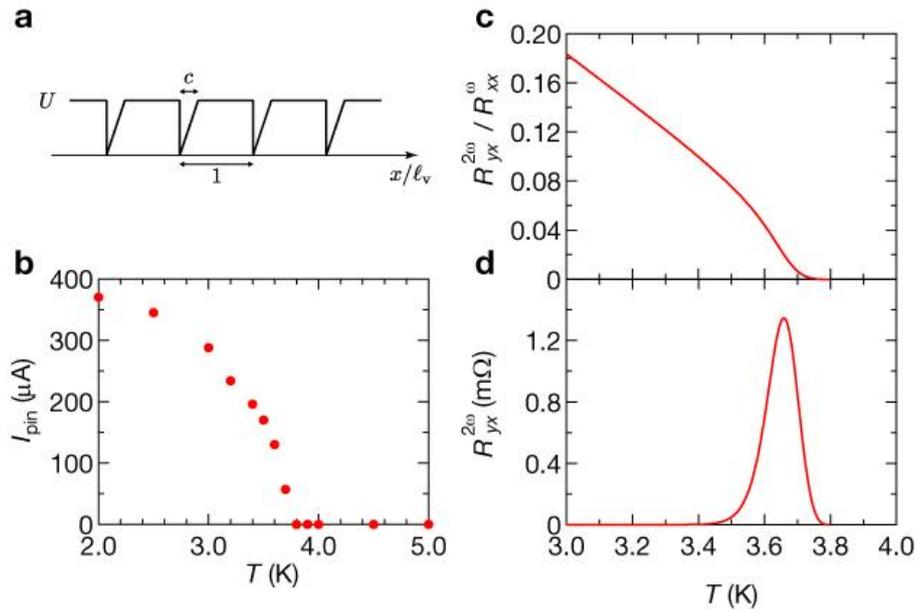

**Figure S4. Theoretically estimated nonlinear transport signals with configuration A.** (**a**) Ratchet potential model used in this paper. (**b**) Temperature dependence of the critical current $I_{pin}$ in sample 6. (**c**) Ratio between nonlinear and linear transport signals near the transition temperature ($T_c \simeq T_{KT} \simeq 3.8$ K). (**d**) Temperature dependence of the nonlinear Hall resistance.

## 2. Comparison with signals in normal state

We compare the transport signals in the superconducting state with those of the normal state[21]. According to Ref. 21, the linear and nonlinear transport signals in the normal state with time-reversal symmetry can be obtained from the Boltzmann transport theory as $j = \sigma_1 E + \sigma_2 E^2$ where

$$\sigma_1 \sim e n_e v_F \frac{e\tau}{p_F}, \qquad \sigma_2 \sim e n_e v_F \left(\frac{e\tau}{p_F}\right)^2 \frac{\tau}{\tilde{\tau}} \quad (S14)$$

Here $v_F$ is the Fermi velocity and $p_F$ is the Fermi momentum, and then the Fermi energy is given by $\varepsilon_F = \frac{1}{2} v_F p_F$. $n_e \sim k_F^2$ is the number density of the normal electrons ($k_F$ is the Fermi wavenumber). $\tau$ and $\tilde{\tau}$ are the scattering times for symmetric and skew scatterings, respectively. From these expressions we may evaluate the resistance ratio $R^{2\omega}/R^\omega$ for the normal state. As for the superconducting state, we have the ratio in Eq. S12 and we take the low-temperature limit for the estimation of its magnitude. The ratio between normal and superconducting states is

$$\frac{(R^{2\omega}/R^\omega)_S}{(R^{2\omega}/R^\omega)_N} \sim \frac{\eta_\perp}{\eta_\parallel} \cdot \frac{\tilde{\tau}}{\tau} \cdot k_F \ell_v \cdot \frac{\varepsilon_F}{k_B T} \quad (S15)$$

where $\eta_\perp/\eta_\parallel$ and $\tau/\tilde{\tau}$ are related to the Hall angles of the vortex and skew scattering of electron, respectively. Assuming that these factors are comparable, the difference in nonlinear transport signals is explained by the two huge factors $k_F \ell_v$ and $\varepsilon_F/k_B T$. Namely, the signal is much enhanced in the superconducting state because the nearly atomic length ($k_F^{-1}$) and Fermi energy are replaced by the much larger characteristic length for the vortices and the temperature, respectively, in the superconducting regime. This enhancement below superconducting transition temperature is consistent with the experimental observation discussed in Fig. 4 of the main text.

## 3. Under magnetic field

In the presence of the external magnetic field, the number of vortices is determined by the total magnetic flux. Then the number densities are given by $n^-(B) = 0$ and $n^+(B) = \frac{B}{\phi_0^*}$. Using Eq. S6 we obtain the transport coefficients as

$$R_{xx}^\omega(B) = \frac{\phi_0^* B L}{\eta_0 W} g_1(\beta U), \qquad R_{xx}^{2\omega}(B) = \frac{(\phi_0^*)^2 B \ell_v L I}{2\eta_0 k_B T W^2} g_2(\beta U) \quad (S16)$$

$$R_{yx}^\omega(B) = \frac{Wr}{L} R_{xx}^\omega(B), \qquad R_{yx}^{2\omega}(B) = \frac{2Wr}{L} R_{xx}^{2\omega}(B) \quad (S17)$$

It is notable that the nonlinear Ohmic signal $R_{xx}^{2\omega}(B)$ is now finite and is not dependent on $r$ originating from the vortex Hall effect. Here again the ratio between linear and nonlinear coefficients is given by the simple quantity

$$\frac{R_{xx}^{2\omega}(B)}{R_{xx}^\omega(B)} = \frac{\phi_0^* \ell_v I}{2 k_B T W} \cdot \frac{g_2(\beta U)}{g_1(\beta U)} \quad (S18)$$

Assuming that the properties of the pinning compared to the zero-field case remain unchanged, the ratio $R^{2\omega}(B)/R^\omega(B)$ in the magnetic field can be larger by the factor $r^{-1} = \eta_\parallel/\eta_\perp$ compared with Eq. S12 for the zero-field case.

## V. Nonlinear anomalous transport in other samples

Nonlinear anomalous Hall effect and the rectification effect have been observed in other samples. In Table S1, we summarized the nonlinear anomalous transport of all the samples we measured. In all samples, directional dependence of the nonlinear anomalous transport is consistent with the symmetry (configurations) and the magnitude of the nonlinear transport are in the same order.

As representatives, we show the current dependence of $R^{2\omega}$ of sample 1 (configuration A, $T$ = 20 K) and sample 2 (configuration B, $T$ = 50 K) in Figs. S5a and S5b, and the current dependence of $R^{2\omega}$ (left) and $R^{\omega}_{xx}$ (right) of sample 6 (configuration A) and sample 7 (configuration B) at $T$ = 2 K in Figs. S5c and S5d. All the data of nonlinear anomalous transport ($R^{2\omega}$) show characteristic selection rules same as that mentioned in the main text (Figs. 2b, 2c, 3a and 3b) or in the previous section (Supplementary Materials III). In Fig. S5e, we plotted $\frac{|E_y^{(2)}|}{\left(E_x^{(1)}\right)^2}$ vs. $(\sigma^{\omega}_{xx})^2$ in sample 1 (green squares) and sample 3 (pink triangles) with configuration A. In both samples, $\frac{|E_y^{(2)}|}{\left(E_x^{(1)}\right)^2}$ shows linear dependence on $(\sigma^{\omega}_{xx})^2$. Black and orange dotted lines show the fitting of $\frac{|E_y^{(2)}|}{\left(E_x^{(1)}\right)^2}$ in sample 1 and 3, respectively, by $\frac{|E_y^{(2)}|}{\left(E_x^{(1)}\right)^2} = \xi(\sigma^{\omega}_{xx})^2 + \eta$, where $\xi$ and $\eta$ are phenomenological fitting parameters. Here, $\xi$ and $\eta$ are estimated as $\xi$ = $2.3 \times 10^{-20}$ m$^3$V$^{-1}\Omega^2$ ($2.2 \times 10^{-20}$ m$^3$V$^{-1}\Omega^2$) and $\eta$ = $-3.2$ μmV$^{-1}$, ($\eta$ = $-30$ μmV$^{-1}$) for sample 1 (sample 3). It is noted that values of $\xi$ in two samples are in similar order, implying that skew scattering induced nonlinear Hall effect is also dominant in these samples.

| Sample No. | configuration | $R_{xx}^{\omega}$ (mΩ) (T = 5 K) | $R_{xx}^{2\omega}$ (mΩ) (normal) | $R_{yx}^{2\omega}$ (mΩ) (normal) | $R_{xx}^{2\omega}$ (mΩ) (SC) | $R_{yx}^{2\omega}$ (mΩ) (SC) |
|---|---|---|---|---|---|---|
| 1 | A | 200 | 0.09 (20 K) | **1.4 (20 K)** | 0.76 (150 μA) | **2.4 (150 μA)** |
| 2 | B | 48 | **0.22 (50 K)** | -0.052 (50 K) | **1.0 (520 μA)** | -0.34 (520 μA) |
| 3 | A | 47 | 0.1 (20 K) | **0.61 (20 K)** | -0.4 (170 μA) | **3.8 (170 μA)** |
| 4 | B | 28 | **1.4 (50 K)** | 0.024 (50 K) | **1.1 (80 μA)** | -0.030 (80 μA) |
| 5 | B | 87 | **0.57 (20 K)** | -0.11 (20 K) | **4.8 (90 μA)** | 1.8 (90 μA) |
| 6 | A | 110 | -0.10 (20 K) | **0.22 (20 K)** | -0.34 (310 μA) | **1.6 (310 μA)** |
| 7 | B | 180 | **0.52 (50 K)** | -0.05 (50 K) | **5.8 (270 μA)** | -0.78 (270 μA) |

**Table S1. Summary of the nonlinear anomalous transport in all samples.** $R_{xx}^{2\omega}$ and $R_{yx}^{2\omega}$ in the normal state is obtained at $I = 3$ mA. $R_{xx}^{2\omega}$ and $R_{yx}^{2\omega}$ in the superconducting state is estimated from the peak of $R_{xx}^{2\omega}$ ($R_{yx}^{2\omega}$) vs. $I$ curves at $T = 2$ K.

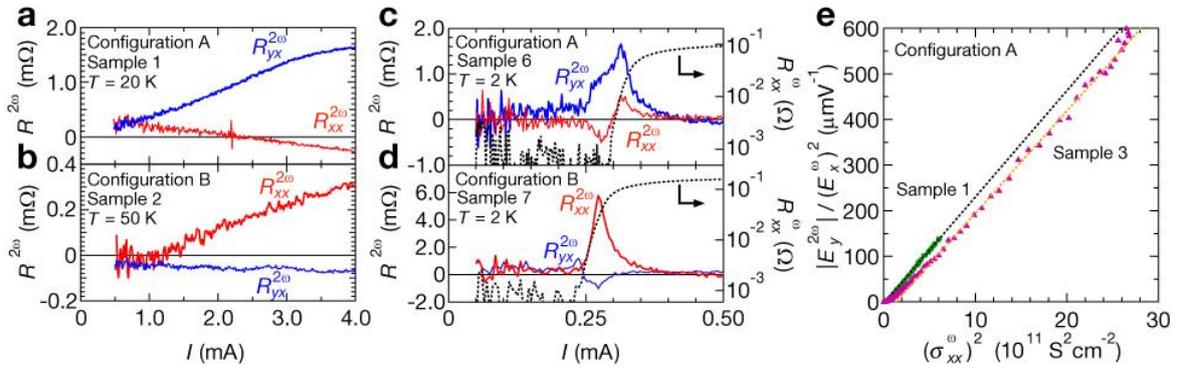

**Figure S5. Nonlinear anomalous transport under time-reversal symmetry in the normal state and superconducting state.** (**a**) and (**b**), Current dependence of the second-harmonic resistance $R^{2\omega}$ in configuration A (sample 1) at $T = 20$ K (a) and configuration B (sample 2) at $T = 50$ K. (**c**) and (**d**), Current dependence of $R^{2\omega}$ (left) and $R_{xx}^{\omega}$ (right) at $T = 2$ K in configuration A (sample 6) (c) and configuration B (sample 7) (d). In Figs. a-d, red and blue

lines indicate longitudinal ($R_{xx}^{2\omega}$) and transversal ($R_{yx}^{2\omega}$) resistance, respectively. (**e**), Normalized second-harmonic response $\frac{|E_y^{(2)}|}{\left(E_x^{(1)}\right)^2}$ as a function of $(\sigma_{xx}^{\omega})^2$ in sample 1 (green squares) and sample 3 (pink triangles). Sample 1 and 3 belong to configuration A. Black and orange dotted lines indicate the linear fitting in sample 1 and 3, respectively.

## VI. Nonlinear superconducting transport under the magnetic field

To achieve the comprehensive understanding of the nonlinear superconducting transport in trigonal PbTaSe$_2$, we measured the second-harmonic resistance under the magnetic field[22]. Figures S6a and S6b show the magnetic field dependence of $R_{xx}^{\omega}$ (Fig. S6a), $R_{xx}^{2\omega}$ (red) and $R_{yx}^{2\omega}$ (blue) (Fig. S6b), in sample 6. In this measurement, the current flows along the zigzag direction (configuration A). We observed clear peak structure in $R_{xx}^{2\omega}$ during the superconducting transition, whose sign is reversed by inverting the magnetic field. Such signals are indiscernible in $R_{yx}^{2\omega}$. Such behavior of nonlinear transport under the magnetic field is called nonreciprocal magnetotransport (or magnetochiral anisotropy), which indicates the rectification effect under broken time-reversal symmetry, and is consistent with previous studies[22,23]. It is important to note that the directional dependence of the second-harmonic signals (i.e., directions along which nonlinear transport appear) are rotated by 90 degree between the nonlinear anomalous transport under time-reversal symmetric condition and nonreciprocal magnetotransport, which are all consistent with the symmetry argument. These results strongly suggest the intrinsic nature of the observed signals. In Figs. S6c and S6d, we also show the temperature dependence of $R_{xx}^{\omega}$ (Fig. S6c), $R_{xx}^{2\omega}$ (red) and $R_{yx}^{2\omega}$ (blue) (Fig. S6d) in sample 6. As discussed in the main text, we observe the large peak in $R_{yx}^{2\omega}$ during the superconducting transition and negligible signals in $R_{xx}^{2\omega}$.

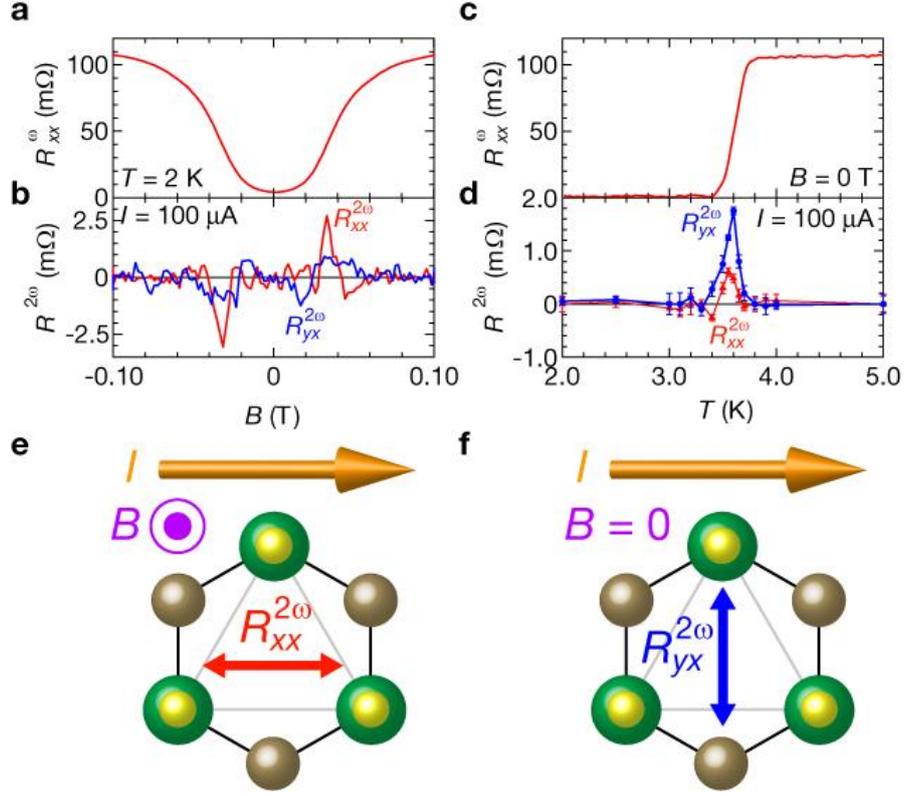

**Figure S6. Nonreciprocal magnetotransport in PbTaSe$_2$.** (**a**, **b**) Magnetic field dependence of the first harmonic longitudinal resistance ($R_{xx}^{\omega}$) and second-harmonic resistance ($R^{2\omega}$) in the longitudinal ($R_{xx}^{2\omega}$) and transverse ($R_{yx}^{2\omega}$) directions at $T = 2$ K and $I = 100$ μA in sample 6. Red (blue) curve in Fig. b shows $R_{xx}^{2\omega}$ ($R_{yx}^{2\omega}$). (**c, d**) Temperature dependence of the first harmonic longitudinal resistance ($R_{xx}^{\omega}$) and second-harmonic resistance ($R^{2\omega}$) in the longitudinal ($R_{xx}^{2\omega}$) and transverse ($R_{yx}^{2\omega}$) directions at $B = 0$ T and $I = 100$ μA in sample 6. Red (blue) curve in Fig. d shows $R_{xx}^{2\omega}$ ($R_{yx}^{2\omega}$). When the magnetic field (~ 0.033 T) is applied, $R_{xx}^{2\omega}$ shows peak structure (nonreciprocal magnetoresistance) while $R_{yx}^{2\omega}$ is dominant (nonlinear anomalous Hall effect discussed in the main text) under the zero magnetic field in the superconducting state. (**e, f**) Schematic image of the nonlinear transport in configuration A under out-of-plane magnetic field (e) and without magnetic field (f). Under (without) magnetic field, the second-harmonic signal is expected in the longitudinal (transverse) direction.

# VII. Comparison between nonlinear superconducting transport with and without magnetic field

We now compare the magnitude of nonlinear transport with and without magnetic field. By combining Eqs. S10, S12, S16 and S18, we can obtain the ratio between $R_{yx}^{2\omega}$ and $R_{xx}^{2\omega}(B)/B$, where $R_{yx}^{2\omega}$ and $R_{xx}^{2\omega}(B)$ are nonlinear transverse resistance without magnetic field and nonlinear longitudinal resistance under magnetic field, respectively, as

$$\frac{R_{yx}^{2\omega}}{R_{xx}^{2\omega}(B)/B} = \frac{2W}{L} \phi_0^* n_v r \quad (S19)$$

To estimate the $r$ value from nonlinear transport measurements, we employed the experimental values at $I = 200$ μA in sample 6. Fig. S7a shows the current dependence of $R_{yx}^{2\omega}$ at $T = 2$ K and $B = 0$ T. The experimental value of $R_{yx}^{2\omega} = 0.20$ mΩ at $I = 200$ μA is extracted as shown by black dashed line in Fig. S7a. Fig. S7b shows the magnetic field dependence of $R_{xx}^{2\omega}$ at $T = 2$ K and $I = 200$ μA. By using the peak amplitude of $R_{xx}^{2\omega}(B) = 0.91$ mΩ and the magnetic field at the peak position = 0.0089 T as an approximation, we estimated the experimental value of $R_{xx}^{2\omega}(B)/B = 0.10$ Ω/T. The slope of $R_{xx}^{2\omega}(B)$ as a function of $B$ is shown by gray dashed line in Fig. S7b. By combining the values obtained above and other parameters, we calculated $r \sim 0.0025$, which is close to the value obtained from linear transport ($r = 0.005$-$0.01$) as shown in Supplementary Material section II.

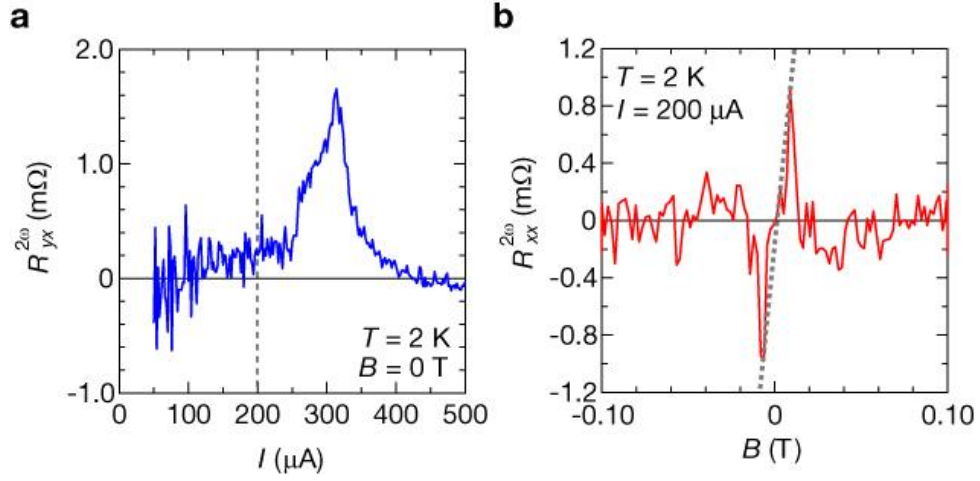

**Figure S7. Comparison between nonlinear transport with and without magnetic field.** (**a**) Current dependence of $R_{yx}^{2\omega}$ at $T = 2$ K and $B = 0$ T in sample 6. Black dashed line means $I = 200$ µA, where we estimated the value of $r$. (**b**) Magnetic field dependence of $R_{xx}^{2\omega}$ at $T = 2$ K and $I = 200$ µA in sample 6. Gray dashed line indicates the slope of $R_{xx}^{2\omega}(B)$ as a function of $B$, which is estimated by using the peak amplitude of $R_{xx}^{2\omega}(B)$ and the magnetic field at the peak position.